# Progress on surface curvature analysis for describing atomization using 2P-LIF images

L. Huang*[1], C.S. Vegad[1], B. Duret[1], J. Reveillon[1], and F. X. Demoulin[1]

[1]University of Rouen Normandy, CNRS CORIA labs, Rouen, France

**Abstract**
To describe atomization completely it is necessary to track the liquid-gas interface morphology at any stage of the atomization process. Typically, instability analysis focuses on generic and simplified morphology: cylindrical jet, liquid sheet, ligament, and droplet to determine their stability and subsequent instability. On the other side sprays composed of spherical droplets are analyzed through their diameter distribution. However, between these situations the liquid-gas interface experiences complex morphology that is more and more accessible through numerical simulation and advanced experimental imagery. To take advantage of this new information and to describe synthetically such data new analyses have been proposed. Here, we aim to analyze complex interface morphology with the surface curvature distribution (SCD) [1] but other possibilities exist [2].
The SCD allows us to describe continuously the destabilization of the initial liquid structure, through complex interfaces such as ligaments, blobs, and liquid sheets until the apparition of the first spherical structures which ultimately become droplets. Beyond the description of the interface, it has been possible to show that a careful analysis of the liquid-gas surface through the SCD allows for determining at the early stage of the atomization process the final characteristics of the spray, even its diameter distribution [3].
In the present work, we are using experimental measurements to assess the characteristics of the spray. With these data, it is possible to observe and describe the atomization process at all stages of the atomization using curvature analysis and image processing techniques.

**Introduction**
In recent years, atomization of the liquid jet has been an increasingly popular topic, especially in the aeronautical engineering and combustion sector. Spray atomization is the transformation of a liquid into a spray of fine particles in a vacuum or a surrounding gas. The breakdown of the liquid into small particles is achieved when compressed air mixes with the liquid. A spray nozzle is used to generate the atomized spray, which passes through an orifice at high pressure and in a controlled manner. This process is widely utilized when distributing material over a cross-section area or generating a liquid surface area over an object. Spray atomization involves creating a condition of high relative velocity between the liquid to be atomized and the surrounding air or gas [4].
Typical atomizer types include electrostatic, ultrasonic, twin-fluid, etc. Twin-fluid atomizers include all nozzle types and may involve two main processes: internal mix and external mix. There are many different kinds of injectors for the atomization process, the most widely used ones are round liquid jets [5] and crossflow liquid jets [6].
There are multiple ways to describe the dense region of the spray experimentally, and each one of them has its advantages and limitations depending largely on the imaging field of interest. For example, Shadowgraph [8] is a line-of-sight imaging technique, due to this effect, the images contain an out-of-focus spray regime. Thus, postprocessing results may also collect unwanted information.

Mie-scattering is one of the laser-illuminated detection techniques, it is often used for the imaging of dense spray. However, this specific method might add unwanted noise (e.g. non-physical spray information) in the final images, hence bringing more difficulties for post-processing of the experimental data. Some possible remedies including structured illumination have been introduced and implemented to solve the issue of blurring effect due to multiple scattering [9,10,11]. There's also another promising technique for imaging dense regions of spray called laser-induced fluorescence (LIF) detection, it can provide high-contrast images, especially for dense spray [12]. This LIF detection method can be applied to a variety of liquid structures with simplified morphology such as liquid sheets, ligaments, or droplets [7].

Nevertheless, among all these candidates, the 2p-LIF (two-photon laser-induced fluorescence) technique proves to capture the information of dense spray at the outlet of the injector at high precision using high contrast images [13,14,15].

The primary break-up of a liquid jet results in the formation of non-spherical droplets and liquid ligaments. Depending on the Weber number, a secondary break-up may occur, resulting in the formation of smaller, spherical droplets. The droplet size distribution (DSD) is measured after the completion of the secondary break-up to obtain a high sphericity validation rate. However, the ligaments and droplets formed during the primary break-up process contain valuable information about the final spray size distribution [3], the reason why we would like to predict the drop size distribution after the secondary breakup using the near-field information.

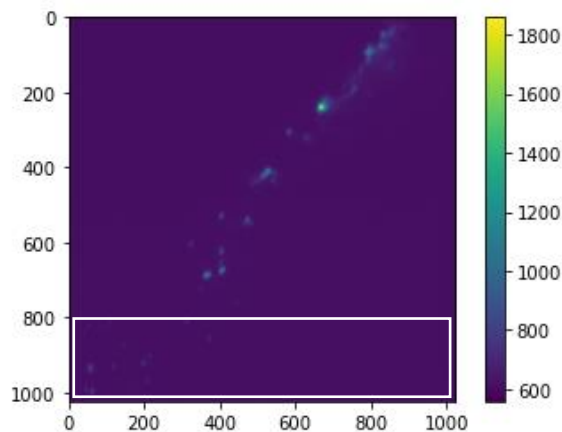

**Figure 1** 2P-LIF image (2.5mm*2.5mm or 1024 pixels*1024 pixels) of the hollow cone spray. The injection starts at an axial distance Z = 0 mm (injector outlet). Considering the symmetry of the spray, only the left half was imaged and shown here. The dispersed region (in the white frame) is used for the collection of the information.

In the current experiment, a pressure swirl atomizer was adopted through the use of a Danfoss commercial nozzle, the experimental data that we focus on are 2p-lif (two-photon laser-induced fluorescence) images represented in Fig. 1, as described in [7,24]. These images were acquired by measuring spray quantities in dense regions of atomized n-heptane conical liquid sheet.

The surface-curvature analysis and postprocessing techniques are used to acquire information in the dense region in Fig. 1. The drop size distribution is obtained from the phase Doppler anemometry measurement [7] in the far field region where most of the spherical droplets are formed after the secondary breakup. The final comparison between the prediction on dense region and far-field measurement shows interesting results, the predicted Sauter

Mean Diameter is found to be in good agreement with the one obtained from measurement, and the predicted drop size distribution follows the close trend of the measured quantity.

**Experimental setup**
In Fig. 2, in the current pressure swirl atomizer, the fuel is firstly injected from three radially positioned cylinders (inlet) on the same horizontal level, each one has a 120-degree spacing from the other one. The mixing of fuel with the airflow inside is guaranteed by the rotation of the internal chamber. From that, the mixed fluid is pushed out through a converging and then slightly diverging orifice and forms a highly atomized hollow cone liquid sheet eventually, as shown in Fig. 3. In the figure, the overall spray topology can be seen in the instantaneous shadowgraphy image, and the half of the spray obtained with other imaging techniques presents the continuous liquid stream to its breakup into non-spherical complex liquid structures.

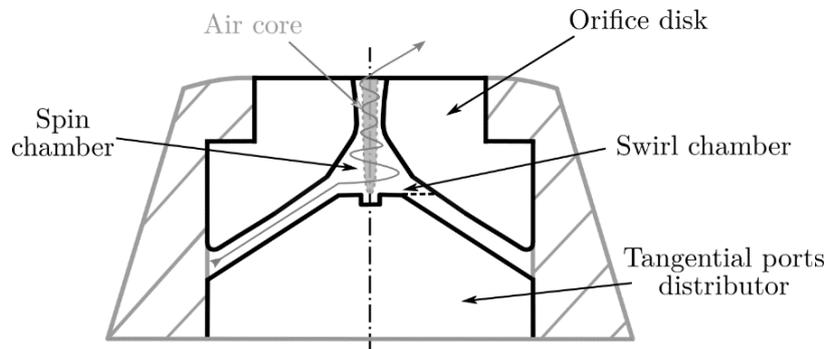

**Figure 2** Sketch of the pressure swirl atomizer. The liquid is injected from the tangential ports and remixed with the air core in the center by the swirl chamber. A hollow cone spray is formed at the orifice.

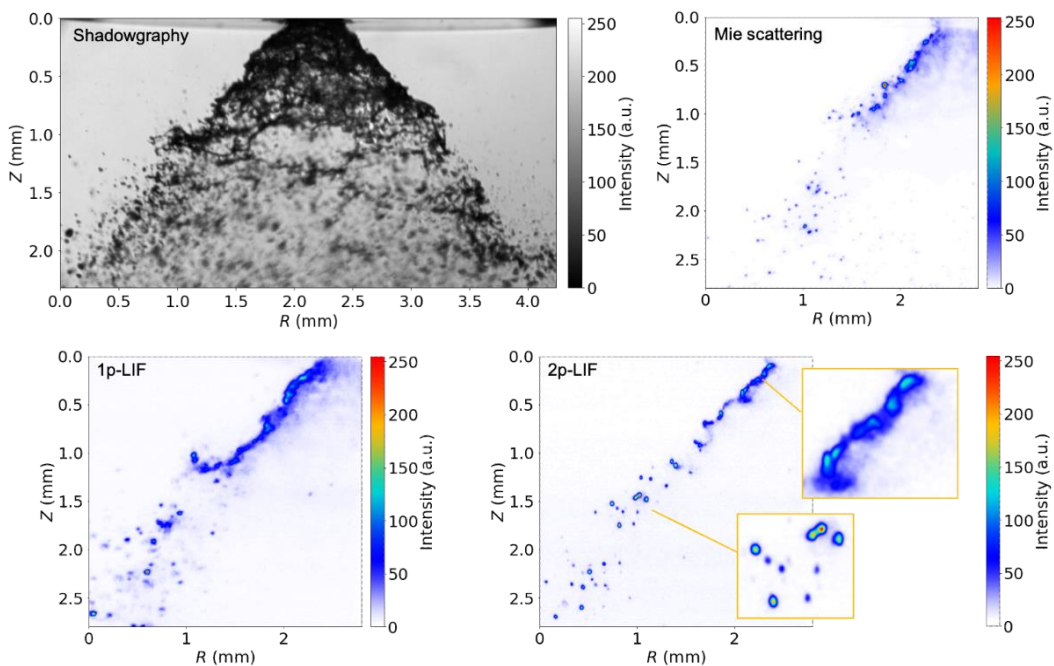

**Figure 3** Shadowgraph image is used to show the general structure of the hollow cone spray. Taking the symmetry property of the spray, only half of the spray is shown using Mie-scattering, 1P-LIF, and 2P-LIF. Details can be found in [7,24].

The pressure swirl atomizer used for the generation of spray is part of the CRSB (CORIA Rouen Spray Burner) system. Its characteristics and technical parameters have been well documented [16,17], and many works [18,19] have been done concerning the imaging and analysis of the secondary breakup region of the hollow cone spray.

**Curvature analysis of image data**

This curvature analysis was originally proposed to study the morphology changes on the liquid-gas interface [1]. Later on, it was applied to numerical simulations of pre-filmer [20] and pressure swirl atomizer [21] to predict the final stage of the atomization process, some other methods such as velocity joint distribution [22] have been proposed, showing promising results. Here a tentative of applying this technique to experimental 2p-lif images will be introduced since it was originally dedicated to numerical simulations.

The most important part of collecting information from two-dimensional images compared to three-dimensional simulations, apart from equations, is the way how we filter the noise and smooth the interface after thresholding the images. These points will be clarified step by step in the form of a workflow, from a raw image to drop size distribution.

Before everything starts, denoising the images is critical to avoid unwanted information interpreted by surface curvature analysis. The first step will be choosing the right strategy to smooth the raw images. Here Gaussian smoothing is used with the radius of the kernel equal to 8 pixels and a standard deviation of 2 pixels. The set of parameters was chosen firstly to smooth the interface of liquid structures in the raw images as illustrated in Fig. 4, the peaks on the interface of liquid are unphysical because of the existence of surface tension. The results of the final prediction depend highly on the way how these peaks are removed since the calculation of curvature is extremely sensitive to the smoothness of the interface and all these peaks will be falsely interpreted as spherical droplets with small diameters and very large curvature values, hence compromising the precision of our surface curvature analysis. On the other hand, the choice of the standard deviation of the Gaussian kernel shouldn't be too large either as it will oversmooth and change the morphology of the liquid structures.

Even though there will be no such problems in simulations with a high resolution where there will be no noise, experiments give us direct access to the study of real complex flow despite some inherent difficulties from the real system that brings possible noise and bias.

After this first step, a proper thresholding of the images needs to be done to define the contour of the liquid structures that are present in the filtered images. Here the threshold value is chosen with the help of the Otsu thresholding technique. Its influence on the proper contour selection of the liquid structures has been explained in [23]. It is interesting to point out that its overestimation or underestimation effect may not seem to be predominant in our postprocessing results compared to its decisive functionality in other fields such as image segmentation.

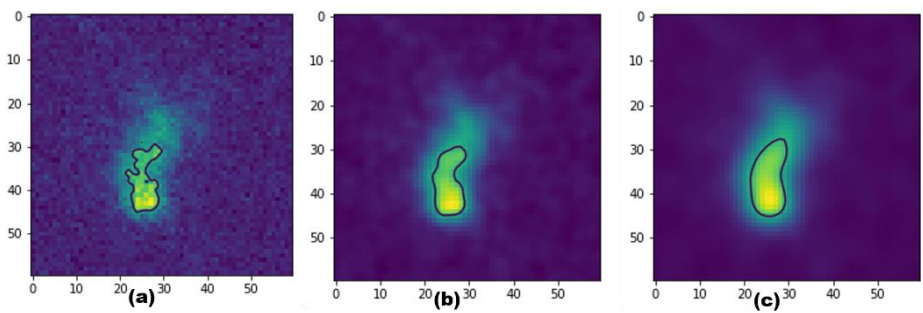

**Figure 4** Images of the sampled liquid structure contoured by the same threshold value. (a) raw image. (b) smoothed by a Gaussian kernel with a standard deviation equal to 1. (c) smoothed by a Gaussian kernel with a standard deviation equal to 2 (adopted version).

The surface curvature analysis begins firstly with the computation of a distance function with the contours of the liquid-gas interface as its zero level set. The thresholding value and the coordinates of the related pixels serve to define where the liquid-gas interface is or where the zero level set of the distance function is, with positive values representing the liquid while negative values representing the gas. The curvature is calculated using the following equation:

$$\kappa = -\nabla \cdot \frac{\nabla \varphi}{|\nabla \varphi|} \tag{1}$$

Moreover, based on the computed distance function, a projected phase indicator (PPI) can be defined using:

$$ppi = (1 + \tanh(\varphi))/2 \tag{2}$$

Where φ is the distance function used for images or phase indicator using the volume-of-fluid method of a simulation. Summarizing the expressions for three-dimensional numerical simulations [22] and two-dimensional images [7,23], the total length or perimeter of the liquid structures over the ppi field of an image or the total surface over the phase indicator using the volume-of-fluid method of a simulation is:

$$L_s = \int_s |\nabla ppi| ds \tag{3}$$

The total surface of the liquid structures over the ppi field of an image or the total volume over the phase indicator using volume-of-fluid method of a simulation is:

$$S_s = \int_s ppi\, ds \tag{4}$$

In three-dimension numerical simulations, a $l_{32}$ defined as the following:

$$l_{32}(v) = 6\frac{V_l}{A_l} = 6\frac{\int_{vol} \alpha\, dv}{\int_{vol} |\nabla \alpha| dv} \tag{5}$$

Where α is the phase indicator. It is used to characterize the general liquid structures and interface before the atomization ends. And a $D_{32}$ is defined as the following:

$$D_{32}(n) = \frac{\sum_i n_i D_i^3}{\sum_i n_i D_i^2} \tag{6}$$

Commonly known as Sauter Mean Diameter, it is used to characterize the drop size distribution after a complete atomization.
In two-dimension images, the alternative expressions have been proposed in the following firstly for $l_{21}$:

$$l_{21}(S) = 4\frac{A_l}{P_l} = 4\frac{\int_S ppi\, ds}{\int_S |\nabla ppi|\, ds} \tag{7}$$

Then for $D_{21}$:

$$D_{21} = \frac{\sum_i n_i D_i^2}{\sum_i n_i D_i} \tag{8}$$

These equations are used to interpret the information of a circle in two dimensions instead of a sphere in three dimensions. Finally, the drop size distribution for images is estimated using the following equations:

$$D_\kappa = \frac{2}{\kappa} \tag{9}$$

$$n_\kappa = \frac{|\nabla ppi|}{\pi D_\kappa} \tag{10}$$

Instead in simulations:

$$D_\kappa = \frac{4}{\kappa} \tag{11}$$

$$n_\kappa = \frac{|\nabla \alpha|}{\pi D_\kappa^2} \tag{12}$$

To sum up, some of the primary parameters between 3D and 2D are listed in the following table 1.

Table 1 Differences between 3D and 2D analysis

| Parameter | 3D simulation | 2D image |
|---|---|---|
| Phase indicator | $\alpha$(vof) | ppi |
| Total volume(3D) surface(2D) | $\int_v \alpha\, dv$ | $\int_S ppi\, ds$ |
| Total surface(3D) length(2D) | $\int_v |\nabla\alpha|\, dv$ | $\int_S |\nabla ppi|\, ds$ |
| Curvature | $-\nabla \cdot \frac{\nabla\alpha}{|\nabla\alpha|}$ | $-\nabla \cdot \frac{\nabla\varphi}{|\nabla\varphi|}$ |

The curvature-related diameter $D_\kappa$ is computed and collected at each cell/pixel location over a certain window using Eq. (9) and is weighted by the number of circular droplets $n_\kappa$ at each location using Eq. (10). The drop size distribution is the histogram of $D_\kappa$ weighted by $n_\kappa$, neglecting the negative and very large diameters at certain locations. A bounded, continuous function is necessary to rescale the distance function between a certain range for the precise calculation of $l_{21}$ in Eq. (7) and $D_{21}$ in Eq. (8). It also serves to smooth the interface for the calculation of the amount of surface and perimeter in Eq. (4) and Eq. (3) at each pixel location.

Here a hyperbolic tangent function is used to help rescale the distance function between -1 and 1, and then it is renormalized between 0 and 1. A test case of the necessity of smoothing the interface is presented in the next section.

A second noise removal process is mandatory for surface curvature analysis. During the experiment, some images are generated containing high-intensity signals, a part of such an image in comparison with a part of a normal image is presented in Fig. 5. General thresholding techniques (e.g. OTSU thresholding) won't be able to remove these signals simply because of their high intensity. These noises contoured by the threshold value, if not filtered, will be treated as normal liquid structures and greatly affect the final prediction.

A simple way to detect these images is to use the distance function itself. For each image where we have already computed the distance function, we choose a small window of the image where no normal signals are present (top left or bottom right) and execute the following command: if the minimum absolute value of the distance function in the region is greater than a certain value (200 pixels in our case), then the image is marked as "clean" and will be used for surface curvature analysis. Such criteria simply filter the images where there's a condensed level of noise (the distance between fake liquid structures is small).

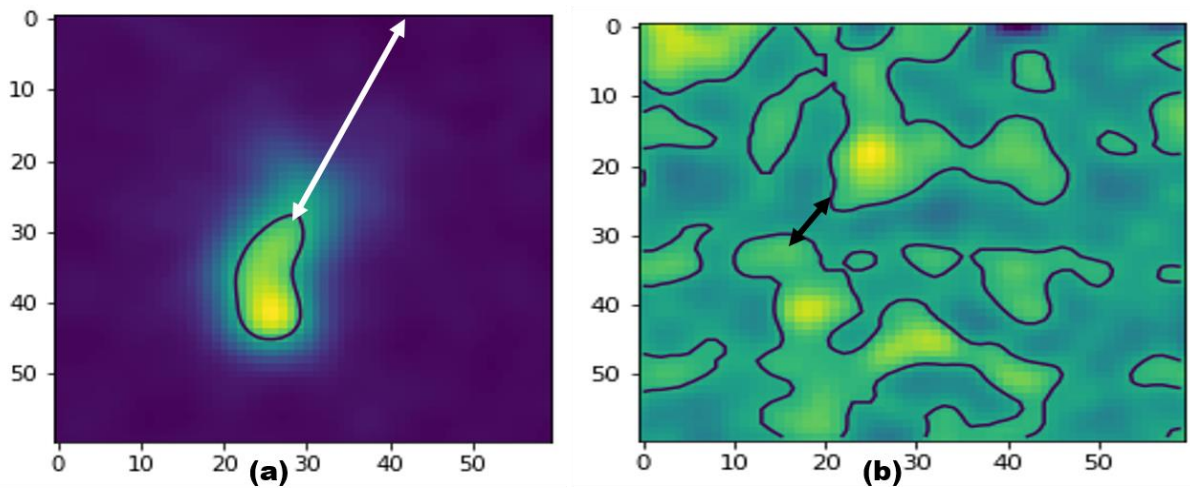

**Figure 5** Part of the images after the first denoising contoured by the same threshold. (a) Normal liquid structure generated during the primary breakup, the white line showing the breakup direction from upstream. (b) Fake liquid structures were detected using the same threshold in the empty zone of the image, the black line showing the breakup direction and the small distance between two "interfaces".

**Testcase**

In Fig. 6 a direct comparison of the drop size distribution with and without interface smoothing is presented. In Fig. 6(e) the predicted drop size distribution with smoothing is calculated over a circle computed using a mathematical equation, it contains almost no noise and represents perfectly the diameter of the circle. In Fig. 6(d), a linearly interpolated image of the same circle is used for the calculation of surface and perimeter, we multiply the distance function map by a diffusion parameter k then binarize the image using two thresholds -1 and 1, and renormalize the map to 0 and 1 in the end. The result is far less stable and precise.

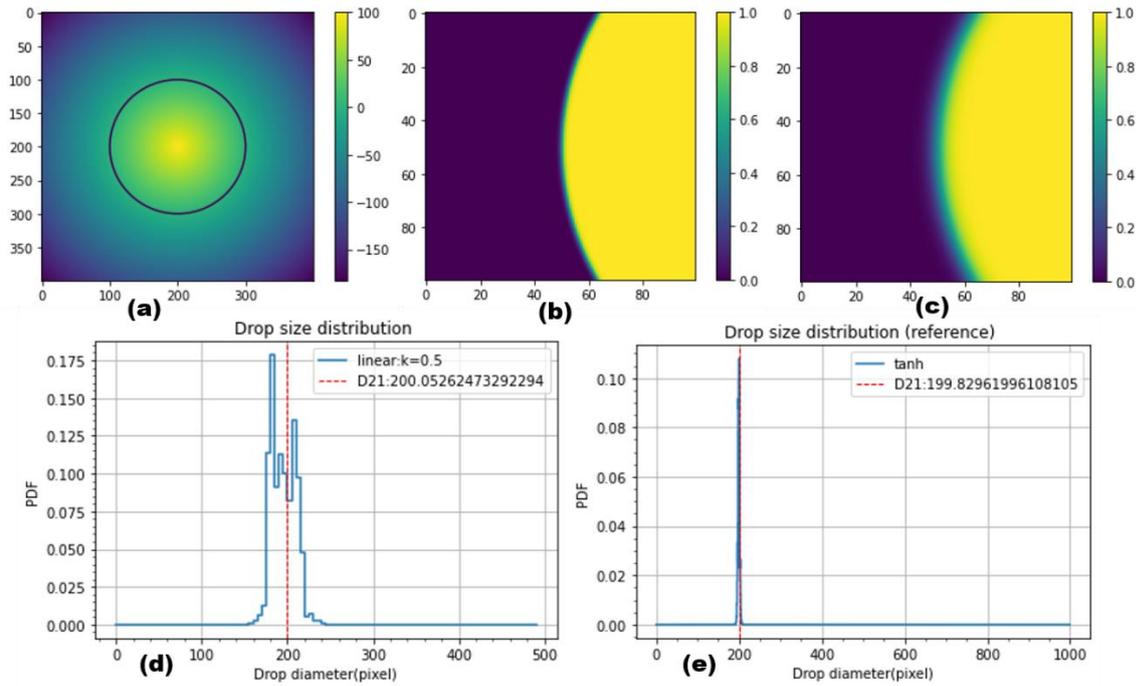

**Figure 6** (a) Distance function of a circle, zero level set (contoured by a black line) represents the circle with a diameter of 200 pixels. (b) A binarized image of part of the circle, the interface is linearly interpolated between 0 and 1. (c) part of the circle using smoothing function, the interface shows non-linear transition. (d) Drop size distribution computed from (b), $D_{21}$ calculated with Eq. (7) and presented by a red dotted line. (e) Drop size distribution computed from (c), $D_{21}$ presented by a red dotted line.

**Results and Discussions**

The surface curvature analysis in three dimensions has been introduced and explained in different works [1,2,3], the two-dimension case is also discussed in [23]. On each pixel location of the image, the following information can be computed: distance function $\varphi$, PPI, curvature using Eq. (1) and (2), estimated diameter and amount of droplets Eq. (9) and Eq. (10). The total amount of perimeter/surface is computed using Eq. (3) and Eq. (4).

The evolution of $D_{21}$ is shown in Fig. 7. When closer to the injector, the $D_{21}$ from Eq. (7) is higher representing the un-atomized liquid bulk, as the region of interest moves further downstream, the primary breakup occurs represented by a range of decreasing values. As the primary breakup ends, the predicted $D_{21}$ converges to the $D_{21}$ from the PDA measurement result showing a good agreement among them, which helps us determine the position of the final window on which we should apply the curvature analysis to get the optimized results.

Figure 8 represents the results after each step of noise removal, the final result after denoising the images and smoothing the interface in Figure 8(c) is in good agreement with the PDA measurement results of spherical droplets 3mm from the injector. The results are computed using the window 2-2.5 mm away from the injector (bottom of the 2P-LIF images). This window, where most of the primary breakups can be observed, corresponds to the place where the computed $D_{21}$ matches the $D_{21}$ from the PDA measurement. The prediction will be less precise if the dense region upstream (e.g. top of the 2P-LIF image) is used where most of the small curvature values that represent big droplets will be collected.

Apart from focusing on the positive curvature values to predict the droplets' sizes, the negative ones have been discarded and should be studied in the future. During the detachment of the liquid structures from the main liquid bulk, the generation of the concave structures carries negative curvature values instead of positive ones like convex liquid droplets. An understanding of the breakup process will be beneficial from the collection of the amount of length or perimeter carried locally by these concave structures. However, some of the processing methods we use (e.g. smoothing in Fig. 4) will change the local morphology from a concave to a convex type bringing more uncertainties into our analysis.

Another major setback is the proper definition of the local threshold value for each window. On the dispersed region (bottom of the 2P-LIF images) where the intensity level is relatively stable, the choice of the threshold for contour detection of the liquid structures is less important. On the continuous region where there's a transition of local intensity level, using one global threshold will not be able to characterize the primary breakup because it will either generate too many fake liquid contours (threshold value too low) or neglect the existence of liquid bulks (threshold value too high). Due to this effect, a link between the operating conditions, experimental setup, and threshold values for each window of the image needs to be established.

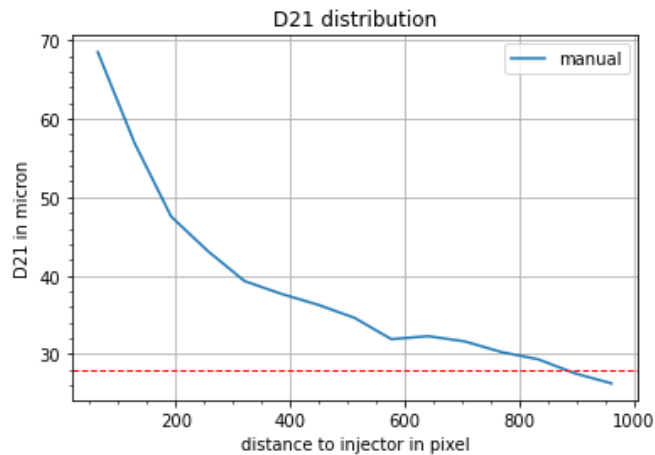

**Figure 7** $D_{21}$ is computed using different windows of the image, each window has an axial size of 1/8 (128 pixels) of the total size of the image (1024 pixels), marching downstream with a step size of 64 pixels each time. The red line represents $D_{21}$ from the PDA measurement result. The threshold values for each window are chosen with the help of the OTSU threshold.

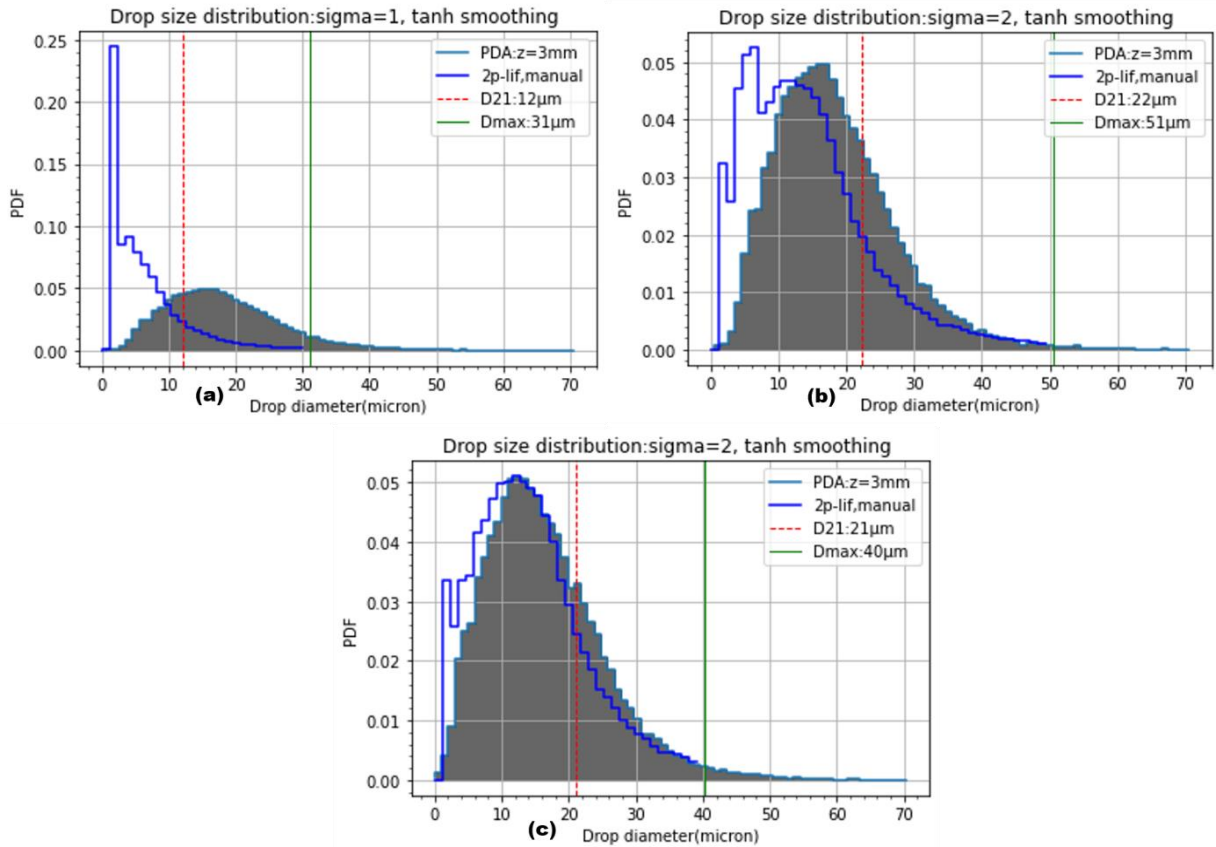

**Figure 8** (a) Drop size distribution computed from partially smoothed images, the red dotted line represents the $D_{21}$ calculated from Eq. (7), the green line represents the Dmax using Eq. (8), diameter range is limited to 0 and 70 microns, all in comparison with the PDA measurement result in the shaded zone. (b) Drop size distribution computed from smoothed images. (c) Drop size distribution computed from a set of smooth, filtered images.

**Conclusion**

The curvature analysis is performed on two-dimensional images as a first tentative. The distance function is used for the computation of curvature and further information. It has been proven that one or several proper denoising techniques play a predominant role in curvature-based prediction. Both the predicted $D_{21}$ and drop size distribution computed using the near field region where primary breakup occurs show good agreement with the related PDA measurement results collected from the far-field region that contains most of the spherical droplet after secondary breakup. This shows that at the early stage of atomization, the liquid structures already carry valuable curvature-related information that helps us better understand the characteristics of the final atomization process.